\documentclass[]{pasj01}

\jyear{2019}

\usepackage{url}

\newcommand{\Ni}{\ensuremath{^{56}\mathrm{Ni}}}

\newcommand{\Msun}{\ensuremath{\mathrm{M}_\odot}}
\newcommand{\Zsun}{\ensuremath{\mathrm{Z}_\odot}}

\newcommand{\Nd}{\ensuremath{\mathrm{N_d}}}
\newcommand{\tint}{\ensuremath{\mathrm{t_{int}}}}

\begin{document} 
\Received{18-Jan-2019}
\Accepted{02-Mar-2019}

\title{Searches for Population III pair-instability supernovae: Predictions for ULTIMATE-Subaru and WFIRST}

\author{Takashi J. \textsc{Moriya}\altaffilmark{1}}
\email{takashi.moriya@nao.ac.jp}

\author{Kenneth C. \textsc{Wong}\altaffilmark{2,1}}

\author{Yusei \textsc{Koyama}\altaffilmark{3}}

\author{Masaomi \textsc{Tanaka}\altaffilmark{4}}

\author{Masamune \textsc{Oguri}\altaffilmark{5,6,2}}

\author{Stefan \textsc{Hilbert}\altaffilmark{7,8}}

\author{Ken'ichi \textsc{Nomoto}\altaffilmark{2}}



\altaffiltext{1}{Division of Theoretical Astronomy, National Astronomical Observatory of Japan, National Institutes of Natural Sciences, 2-21-1 Osawa, Mitaka, Tokyo 181-8588, Japan}

\altaffiltext{2}{Kavli Institute for the Physics and Mathematics of the Universe (WPI), The University of Tokyo Institutes for Advanced Study, The University of Tokyo, 5-1-5 Kashiwanoha, Kashiwa, Chiba 277-8583, Japan}

\altaffiltext{3}{Subaru Telescope, National Astronomical Observatory of Japan, National Institutes of Natural Sciences, 650 North A'ohoku Place, Hilo, HI 96720, USA}

\altaffiltext{4}{Astronomical Institute, Tohoku University, 6-3 Aramaki Aza-Aoba, Aoba-ku, Sendai 980-8578, Japan}

\altaffiltext{5}{Research Center for the Early Universe, Graduate School of Science, The University of Tokyo, 7-3-1 Hongo, Bunkyo, Tokyo 113-0033, Japan}

\altaffiltext{6}{Department of Physics, Graduate School of Science, The University of Tokyo, 7-3-1 Hongo, Bunkyo, Tokyo 113-0033, Japan}

\altaffiltext{7}{Exzellenzcluster Universe, Boltzmannstr. 2, D-85748 Garching, Germany}

\altaffiltext{8}{Ludwig-Maximilians-Universit\"at, Universit\"ats-Sternwarte, Scheinerstr. 1, D-81679 M\"unchen, Germany}



\maketitle

\begin{abstract}
ULTIMATE-Subaru (Ultra-wide Laser Tomographic Imager and MOS with AO for Transcendent Exploration on Subaru) and WFIRST (Wide Field Infra-Red Survey Telescope) are the next generation near-infrared instruments that have a large field-of-view. They allow us to conduct deep and wide transient surveys in near-infrared. Such a near-infrared transient survey enables us to find very distant supernovae that are redshifted to the near-infrared wavelengths. We have performed the mock transient surveys with ULTIMATE-Subaru and WFIRST to investigate their ability to discover Population~III pair-instability supernovae. We found that a 5-year 1~$\mathrm{deg^2}$ $K$-band transient survey with the point-source limiting magnitude of 26.5~mag with ULTIMATE-Subaru may find about 2 Population~III pair-instability supernovae beyond the redshift of 6. A 5-year 10~$\mathrm{deg^2}$ survey with WFIRST reaching 26.5~mag in the $F184$ band may find about 7 Population~III pair-instability supernovae beyond the redshift of 6. We also find that the expected numbers of the Population~III pair-instability supernova detections increase about a factor of 2 if the near-infrared transient surveys are performed towards clusters of galaxies. Other supernovae such as Population~II pair-instability supernovae would also be detected in the same survey. This study demonstrates that the future wide-field near-infrared instruments allow us to investigate the explosions of the first generation supernovae by performing the deep and wide near-infrared transient surveys.
\end{abstract}

\section{Introduction}
The last decade encountered the great success of optical transient surveys. Many optical transient surveys such as PTF \citep{law2009ptf}, Pan-STARRS \citep{chambers2016panstarrs}, SkyMapper \citep{keller2007skymapper}, KISS \citep{morokuma2014kiss}, and HSC (\cite{tanaka2016hscrapidrise}, \cite{moriya2018shizuca}, Yasuda et al. submitted) have shown that stellar deaths have much more varieties than expected before. One of the most fascinating discoveries from these optical transient surveys are the discovery of very luminous supernovae (SNe) called superluminous supernovae (SLSNe, \cite{quimby2011slsn}; see \cite{moriya2018slsnreview} for a review). They are often more than 10 times brighter than canonical SNe. Shortly after their discovery, it has been speculated that their huge luminosities could be due to massive production of radioactive \Ni\ and, therefore, SLSNe might be pair-instability SNe (PISNe) (e.g., \cite{smith2007sn2006gyearly,gal-yam2009sn2007bi}). PISNe are theoretically predicted explosions of very massive stars \citep{rakavy1967pisn,barkat1967pisn}. PISNe could lead to the production of more than 10~\Msun\ of \Ni\ \citep{heger2002popiii}, which is generally required to explain the huge luminosity of SLSNe.

Although PISNe were originally suggested to be a probable origin of SLSNe, it turned out that SLSNe generally show different observational properties than those expected from PISNe (e.g., \cite{kasen2011pisn,dessart2012magslsn,whalen2014pisn,jerkstrand2016pisnnebular,tolstov2017gaia} but see also \cite{kozyreva2016rapidpisn}). PISNe are a natural consequence of very massive stars that have the helium core masses between $\sim 70~\Msun$ and $\sim 140~\Msun$ \citep{langer2012review}, and we would observe them if massive stars with such massive cores exist. Unfortunately, it is very hard to keep the cores massive enough to explode as PISNe until their deaths in the solar metallicity environment (e.g., \cite{yoshida2014massiveic}, but see also \cite{georgy2017solarpisn}) and PISNe are suggested to occur when the metallicity is below one third of the solar metallicity \citep{langer2007pisn}. The most promising stars to explode as PISNe are the first stars, or Population~III (Pop~III) stars. Pop~III stars do not suffer from wind mass loss and they can grow massive enough cores to explode as PISNe when their zero-age main-sequence (ZAMS) masses are between $\sim 140~\Msun$ and $\sim 260~\Msun$ \citep{heger2002popiii,umeda2002popiii}. Pop~III stars are also predicted to be dominated by massive stars including those in the PISN mass range (e.g., \cite{hirano2015popiii}). Therefore, many Pop~III stars are likely to explode as PISNe.

Pop~III star formation could last until $z\sim 6$ (e.g., \cite{desouza2014pisnrate}) and, therefore, we need to reach at least $z\sim 6$ to look for Pop~III PISNe. It is inevitable to perform transient surveys in near infrared (NIR) to search for transients at such high redshifts (e.g., \cite{scannapieco2005pisn,tanaka2013nirslsndetec,whalen2013pisn,desouza2013jwst,hartwig2018jwstpisn}). Only a handful of NIR transient surveys are conducted in the last decade (e.g., \cite{mattila2012nirsn,kool2018sunbird,kasliwal2017spirits}). However, many NIR wide field imaging instruments are currently planned and NIR transient surveys would be the frontier of transient surveys in the coming decade (e.g., \cite{inserra2017euclid,hounsell2017wfirstia}). Among them, ULTIMATE-Subaru (Ultra-wide Laser Tomographic Imager and MOS with AO for Transcendent Exploration\footnote{\url{https://www.naoj.org/Projects/newdev/ngao/}}) and WFIRST (Wide-Field InfraRed Survey Telescope, \cite{spergel2015wfirst}) have a wide-field NIR imaging facility that is suitable to perform deep and wide NIR transient surveys. In this paper, we perform mock NIR transient surveys with ULTIMATE-Subaru and WFIRST to search for Pop~III PISNe at $z\gtrsim 6$ and study optimal survey strategies to detect them. 
We also take the effect of the luminosity amplifications due to gravitational lensing into account (cf. \cite{whalen2013lensing}).
 In this paper, we focus on survey strategies and observational methods to discover Pop~III PISNe and we briefly show the effect of the gravitational lensing in this paper. The details on how gravitational lensing could affect the survey results, as well as the details on how we estimate the gravitational lensing effect, are discussed in an accompanying paper by \citet{wong2019}.

The rest of this paper is organized as follows. We first present our method of mock observations in Section~\ref{sec:setup}. We present our results of mock observations and discuss the optimal strategy to look for Pop~III PISNe at $z\gtrsim 6$ in Section~\ref{sec:results}. We have general discussion of our results in Section~\ref{sec:discussion} and conclude this paper in Section~\ref{sec:conclusions}. Throughout this paper, we adopt the standard $\Lambda$CDM cosmology with $H_0=70~\mathrm{km~s^{-1}~Mpc^{-1}}$, $\Omega_\Lambda = 0.7$, and $\Omega_M = 0.3$. All the magnitudes presented in this paper are in the AB magnitude system.

\begin{figure}
 \begin{center}
  \includegraphics[width=8cm]{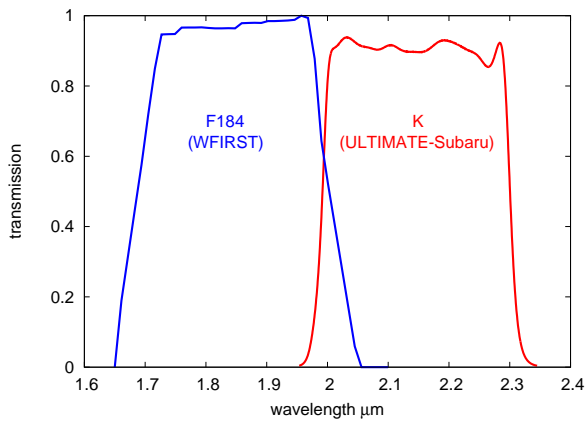} 
 \end{center}
\caption{
Filters used in our SN survey simulations. The $K$-band filter assumed for ULTIMATE-Subaru is that currently used by Subaru/MOIRCS. The $F184$ filter is the reddest filter currently planned by WFIRST. 
}\label{fig:filters}
\end{figure}

\section{Set-up of mock observations}\label{sec:setup}
\subsection{Instruments}
We first introduce ULTIMATE-Subaru and WFIRST. The two instruments are complimentary to each other in terms of the wavelength coverage and the field-of-view (FoV).

\subsubsection{ULTIMATE-Subaru}
ULTIMATE-Subaru is the next generation wide-field AO system with a wide-field NIR instrument. It is currently planned to have the NIR imager with the FoV of $14'$ x $14'$ ($0.054~\mathrm{deg^2}$). It requires 18.4 pointings to cover $1~\mathrm{deg^2}$, for example.

The details of the instrument are still under discussion. The reddest band currently planned is the $K$ band. We expect that the instrument specification such as filters and exposure time of ULTIMATE-Subaru will be similar to those of Multi-Object Infrared Camera and Spectrograph (MOIRCS) currently on Subaru \citep{suzuki2008moircs,ichikawa2006moircs}. We adopt the $K$ band filter of MOIRCS in this study (Fig.~\ref{fig:filters}). According to the exposure time calculator for MOIRCS\footnote{\url{https://www.naoj.org/cgi-bin/img_etc.cgi}}, it takes 2.6 hours to reach the limiting magnitude (signal-to-noise ratio of 5 for a point source throughout this paper) of 26.0~mag in the $K$ band with a standard condition ($0.2"$ seeing with $0.3"$ aperture). With the best expected condition, it is possible to reach $0.15"$ seeing. Then, the limiting magnitude of 26.5~mag in the $K$ band can be reached in 4.6~hours.

\subsubsection{WFIRST}
WFIRST is the next generation space telescope with the wide-field imager that has the FoV of $0.28~\mathrm{deg^2}$ \citep{spergel2015wfirst}. 3.6 pointings are required to cover $1~\mathrm{deg^2}$.

The reddest filter for the imager currently planned is the $F184$ filter (Fig.~\ref{fig:filters}) and we adopt it in this study. The deepest SN survey currently planned with WFIRST has the limiting magnitude of about 26.5~mag. It takes 0.5~hours to reach 26.5~mag in $F184$. The limiting magnitude of 26.0~mag can be reached in 0.2~hours. We also perform a simulation with the limiting magnitude of 27.0~mag, which can be reached in 1.3~hours.

\begin{figure*}
 \begin{center}
  \includegraphics[width=8cm]{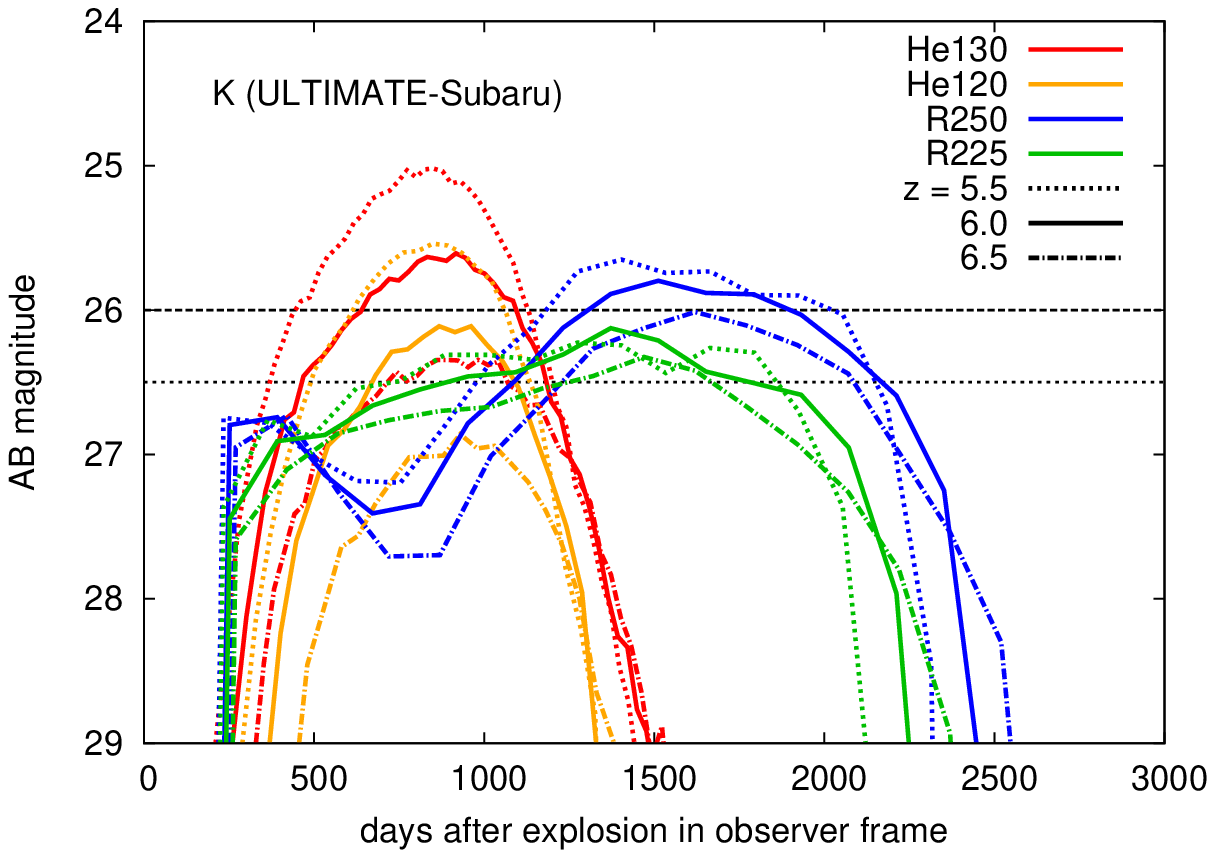} 
  \includegraphics[width=8cm]{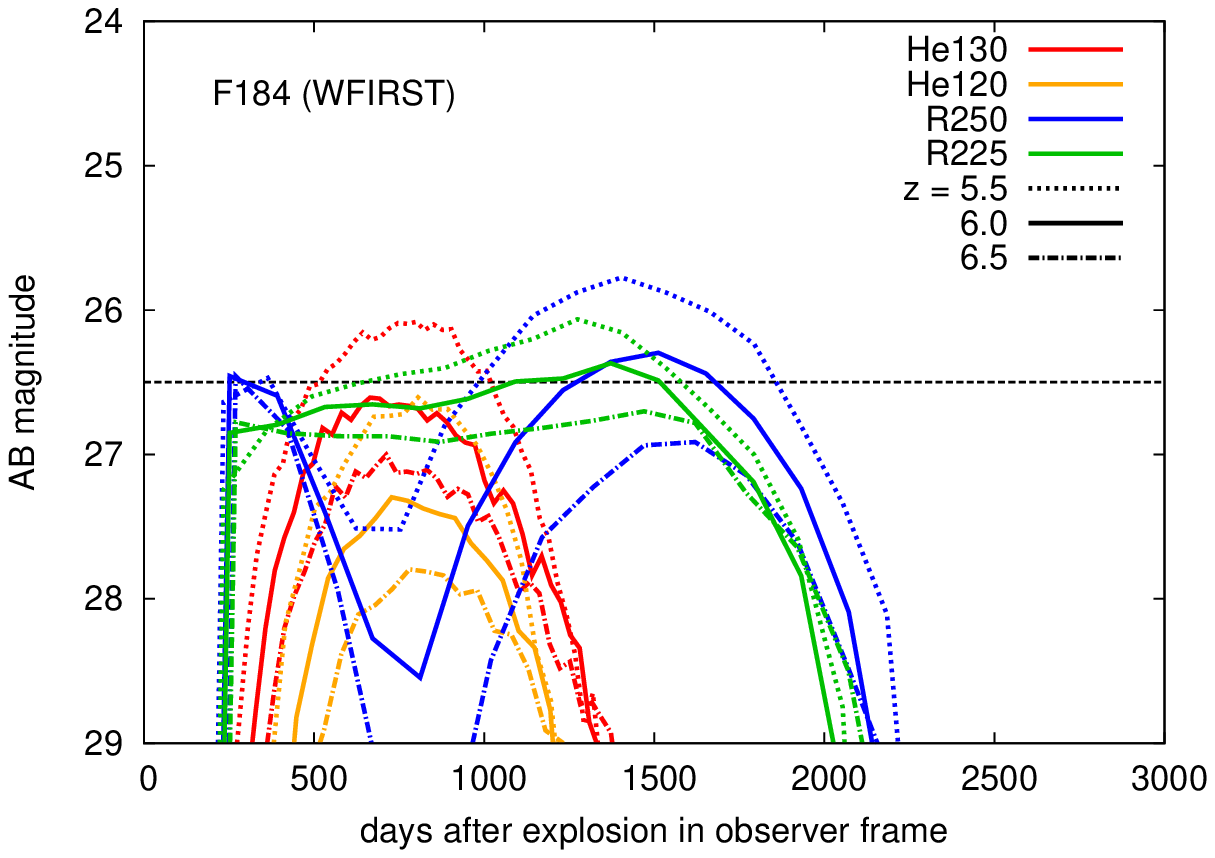}   
 \end{center}
\caption{
Examples of PISN LCs adopted in our mock observation simulations. Left panel shows the $K$-band LCs expected for ULTIMATE-Subaru for which we assume the 26.0~mag and 26.5~mag limits (horizontal lines) in our survey simulations. Right panel shows the $F184$-band LCs for WFIRST for which we assume the 26.5~mag limit (horizontal line) in the deepest survey simulations.
}\label{fig:pisnlcs}
\end{figure*}

\subsection{Survey strategy}\label{sec:strategy}
We perform our mock observation simulations with several different survey strategies. We fix our survey period to be 5~years. We only use one filter in our mock transient surveys, $K$ for ULTIMATE-Subaru and $F184$ for WFIRST.

The survey field is assumed to be observed repeatedly with an interval of \tint. We also set the minimum number of detections \Nd\ for a transient to be regarded as a discovery. For example, if we assume $\Nd=3$, PISNe detected more than 3 times are regarded as PISN discoveries. In this case, PISNe detected only two times are ignored.

For the ULTIMATE-Subaru survey, we assume two different limiting magnitudes, 26.5~mag and 26.0~mag in the $K$ band. It requires about 70~hours to cover $1~\mathrm{deg^2}$ with 26.5~mag for one epoch. Assuming $\tint=180~\mathrm{days}$, the 5-year $1~\mathrm{deg^2}$ transient survey with the 26.5~mag limit takes 860~hours in total. If we adopt 26.0~mag as the survey limiting magnitude, it takes 48~hours to cover $1~\mathrm{deg^2}$. With $\tint=180~\mathrm{days}$ for 5 years, 480~hours and 960~hours are required to conduct the $1~\mathrm{deg^2}$ and $2~\mathrm{deg^2}$ surveys, respectively.

WFIRST requires 2~hours and 0.75~hours to cover $1~\mathrm{deg^2}$ with the limiting magnitude of 26.5~mag and 26.0~mag, respectively. For instance, the transient survey with $\tint=180~\mathrm{days}$ covering $10~\mathrm{deg^2}$ in 5 years takes 200~hours (26.5~mag limit) and 75~hours (26.0~mag limit).

\subsection{PISN properties}
\subsubsection{Light curves}
We adopt PISN light curves (LCs) predicted by \citet{kasen2011pisn} for our observation simulations. The LCs are numerically obtained from the PISN progenitors of \citet{heger2002popiii}. The peak luminosity of the R250 and He130 models (see the next paragraph for the description of the models) exceeds $K=26.0~\mathrm{mag}$ at $z = 6$ (Fig.~\ref{fig:pisnlcs}). The R225 and He120 models are brighter than 26.5~mag in the $K$ band at $z=6$ (Fig.~\ref{fig:pisnlcs}). Thus, the R225 and He120 explosions at $z\gtrsim 6$ can only be observed when we conduct the transient surveys with the limiting magnitude of 26.5~mag in the $K$ band. If we use the $F184$ filter, only R250 and R225 are brighter than $26.5~\mathrm{mag}$ at the peak at $z\gtrsim 6$ (Fig.~\ref{fig:pisnlcs}). For reference, the peak magnitude of the R250 model in the $H$ band of WFIRST \citep{spergel2015wfirst} is around 27.1~mag and it is much more efficient to conduct a survey in the $F184$ band. The $F184$ band magnitudes become significantly faint in He130 and He120 models because the helium core models are redder than the red supergiant models. For example, the synthetic spectra of R250 at around the maximum luminosity peak at $\sim 2500$~\AA, while the helium core models peak at $\sim 3500$~\AA\ \citep{kasen2011pisn}. This difference likely originates from the difference in Fe-group absorption. The red supergiant models have the photosphere in their hydrogen-rich envelope and it is far from the central region where Fe-group elements locate. However, the helium core models have the photosphere at or close to where the \Ni\ is synthesized and they are more affected by absorption by the Fe-group elements. 

R250 and R225 are the red supergiant (RSG) PISN progenitors with the ZAMS mass of 250~\Msun\ and 225~\Msun, respectively. They have the initial metallicity of $10^{-4}~\Zsun$. The masses at the time of the explosion are $236~\Msun$ (R250) and $200~\Msun$ (R225) because of slight mass loss. A similar amount of mass loss is found even in the Pop~III RSG PISN progenitors in \citet{moriya2015langer}. Therefore, the RSG PISN LC properties are not likely to be much different from those of the zero-matellicity progenitors. He130 and He120 are the hydrogen-free PISN progenitors with the initial helium core masses of 130~\Msun\ and 120~\Msun, respectively. It is the zero-metallicity (Pop~III) models and the initial masses are kept until the time of the explosions without mass loss.

\subsubsection{Rates}\label{sec:rates}
\begin{figure}
 \begin{center}
  \includegraphics[width=8cm]{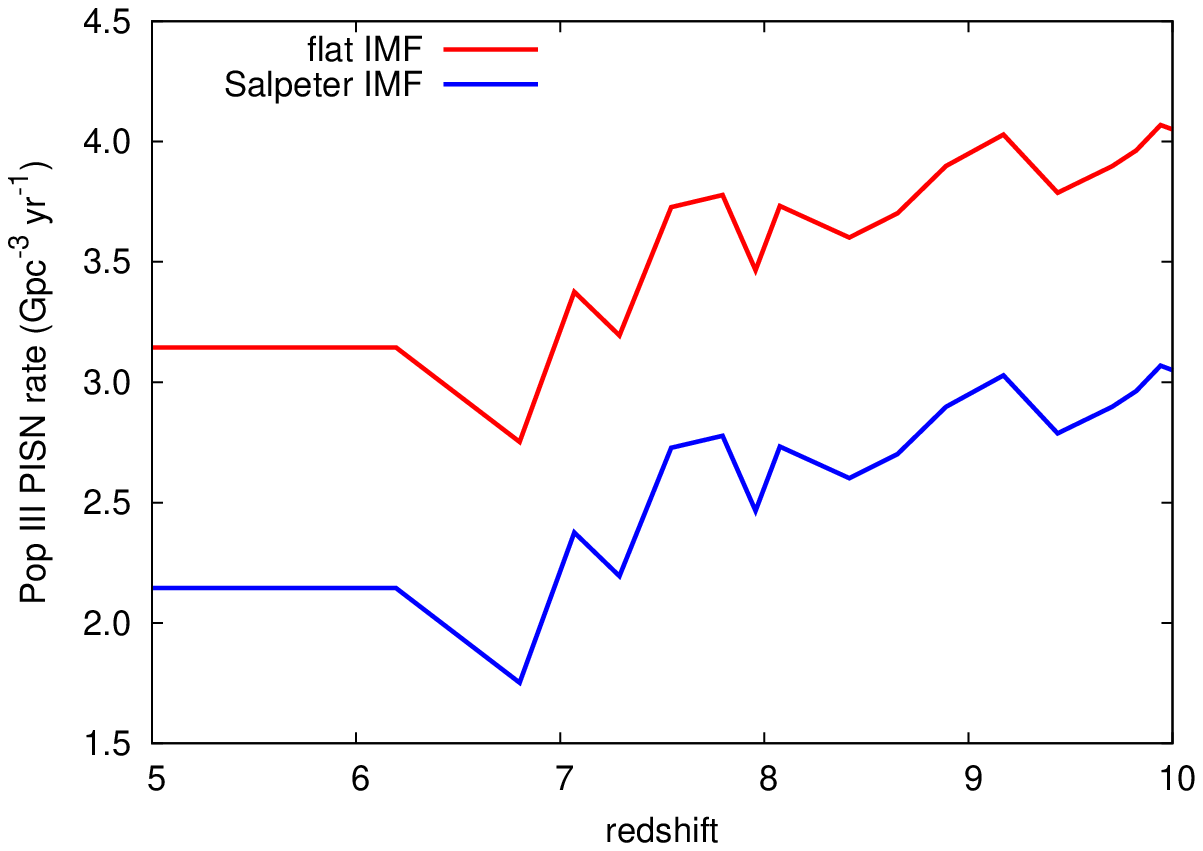} 
 \end{center}
\caption{Pop~III PISN rate estimated by \citet{desouza2014pisnrate}.}\label{fig:pisnrate}
\end{figure}

We adopt a PISN rate estimated by \citet{desouza2014pisnrate} in our mock observations (Fig.~\ref{fig:pisnrate}). \citet{desouza2014pisnrate} estimated the Pop~III PISN rate based on the cosmological simulation of \citet{johnson2013cosmosim}. We take the Pop~III PISN rates estimated from their high SFR estimates (SFR10 in \cite{desouza2014pisnrate}). \citet{desouza2014pisnrate} estimated Pop~III PISN rate above $z=6$. We also present the PISN discovery number estimates at $5<z<6$ by assuming that the PISN rate at $5<z<6$ is the same as that at $z=6$.

There exist many works estimating the Pop~III SFRs. Fig.~2 in \citet{desouza2014pisnrate} compares several predicted Pop~III SFRs. The SFR10 model we adopt has a relatively high SFR prediction among others but other predictions stay roughly within a factor of 10 compared to the SFR10 estimate.

\citet{desouza2014pisnrate} assumed two initial mass functions (IMFs) to estimate the Pop~III PISN rate. One is the Salpeter IMF and the other is the flat IMF. The cosmological simulations of the first stars indicate that the IMF is close to the flat one in Pop~III stars \citep{hirano2015popiii}.

The ZAMS mass range for Pop~III PISNe is roughly between $140~\Msun$ and $260~\Msun$ \citep{heger2002popiii}. Only the R225 (ZAMS mass of 225~\Msun) and R250 (ZAMS mass of 250~\Msun) models in Pop~III RSG PISN models of \citet{kasen2011pisn} become brighter than $K=26.5~\mathrm{mag}$ and $F184 = 26.5~\mathrm{mag}$ when they appear at $z\gtrsim 6$. The R200 model (ZAMS mass of 200~\Msun) does not become bright enough. \citet{kasen2011pisn} does not provide RSG PISN LCs between 200~\Msun\ and 225~\Msun\ and we cannot tell the exact minimum ZAMS mass to be brighter than 26.5~mag. In this study, we assume that RSG Pop~III PISNe whose ZAMS mass is above 215~\Msun\ become bright enough to be observable at $z\sim6$ with the $K=26.5~\mathrm{mag}$ limit.

If we assume the $K=26.0~\mathrm{mag}$ limit, the R225 model at $z\gtrsim 6$ cannot be observed and only R250 is observable. Similarly, \citet{kasen2011pisn} does not provide RSG PISN LCs between 225~\Msun\ and 250~\Msun\ and we cannot tell the exact minimum ZAMS mass to be bright enough to observe with the limit. We assume that RSG Pop~III PISNe whose ZAMS mass is above 240~\Msun\ become bright enough to be observed at $z\sim6$ with the $K=26.0~\mathrm{mag}$ limit. We assume that the R250 LC is the representative LC of Pop~III PISNe whose ZAMS masses are between 240~\Msun\ and 260~\Msun\ and the R225 LC is the representative LC of Pop~III PISNe whose ZAMS masses are between 215~\Msun\ and 240~\Msun.

If we assume the flat IMF, the fraction of Pop~III PISNe with the ZAMS mass between 240~\Msun\ and 260~\Msun\ is 17\%. In our mock observations, we thus assume that the R250 PISNe explode with the 17\% rate of the total Pop~III PISN rate obtained by \citet{desouza2014pisnrate} in the case of the flat IMF. Similarly, 21\% of the Pop~III PISN rate comes from between 215~\Msun\ and 240~\Msun\ with the flat IMF and we assume that the R225 PISNe occupies 21\% of the Pop~III IMF in case of the flat IMF. On the other hand, when we assume the Salpeter IMF, the PISN fraction in the ZAMS mass between $240~\Msun$ and $260~\Msun$ becomes 8.7\% and the PISN fraction in the ZAMS mass between $215~\Msun$ and $240~\Msun$ becomes 14\%. Therefore, we assume that the R250 model and the R225 model explode with the 8.7\% and 14\%, respectively, of the total PISN rate estimated by \citet{desouza2014pisnrate} in the case of the Salpeter IMF.

The bare helium core PISN models He120 and He130 are observable at $z\gtrsim 6$ with our deep surveys (Fig.~\ref{fig:pisnlcs}). However, it is not obvious from which ZAMS masses these massive bare helium core progenitors originate (see Section~\ref{sec:heliumcore} for discussion). Therefore, we perform mock observations only considering Pop~III RSG PISN progenitors. We discuss the possible effect of helium core Pop~III PISN progenitors in Section~\ref{sec:discussion}.

The blue supergiant (BSG) Pop~III PISNe are not observable at $z\gtrsim 6$ with the $K=26.5~\mathrm{mag}$ limit and a $F184=26.5~\mathrm{mag}$ limit. Many stellar evolution models predict that Pop~III SLSNe with hydrogen explode as RSGs, not BSGs (e.g., \cite{yoon2012popiii,moriya2015langer}). Therefore, we ignore Pop~III BSG PISNe in this study.

\begin{longtable}{ccccccccccc}
  \caption{Numbers of Pop~III RSG PISN discoveries for the 5-year survey without the gravitational lensing effect.}\label{tab:pisnnum}
  \hline
 band & FoV & limit & \tint & \Nd & \multicolumn{3}{c}{flat IMF} & \multicolumn{3}{c}{Salpeter IMF}  \\ 
  & $\mathrm{deg^2}$ & mag & days & & $z>5$ & $z>6$ & $z>7$ & $z>5$ & $z>6$ & $z>7$  \\ 
\endfirsthead
\endhead
  \hline
\endfoot
  \hline
  \multicolumn{10}{l}{$^{a}$ULTIMATE-Subaru, $^{b}$WFIRST}
\endlastfoot
  \hline
$K^a$ & 1 & 26.5 & 180 & 2 & $7.9\pm0.4$ & $2.4\pm0.2$ & $0.06\pm0.03$ & $0.4\pm0.1$  & $0.13\pm0.05$ & $0.005 \pm 0.002$ \\
      & 1 & 26.5 & 180 & 3 & $6.4\pm0.5$ & $1.8\pm0.3$ & $0.008\pm 0.004$ & $0.35\pm0.09$ & $0.09\pm0.04$ & $0$ \\
      & 2 & 26.0 & 180 & 2 & $4.9 \pm 0.4$ & $1.0\pm0.2$ & $0$ & $0.24\pm0.06$ & $0.04\pm0.02$ & $0$ \\
      & 2 & 26.0 & 180 & 3 & $3.4\pm0.2$ & $0.3\pm0.1$ & $0$ & $0.3\pm0.1$ & $0.006\pm0.04$ & $0$ \\
      & 2 & 26.0 & 90  & 3 & $5.0\pm0.4$ & $1.2\pm0.2$ & $0$ & $0.24\pm0.06$ & $0.04\pm0.04$ & $0$ \\
$F184^b$ & 1  & 27.0 & 180 & 2 & $10.6\pm 0.3$ & $4.4 \pm 0.2$ & $1.10\pm 0.08$ & $0.60\pm0.05$ & $0.24\pm0.03$ & $0.06\pm0.01$ \\
         & 10 & 26.5 & 180 & 2 & $57\pm2$ & $6.8\pm0.7$ & $0$ & $3.2\pm0.4$ & $0.3\pm0.1$ & $0$ \\
         & 10 & 26.0 & 180 & 2 & $13.2\pm 0.6$ & $0$ & $0$ & $0.7\pm0.1$ & $0$ & $0$ \\
\end{longtable}

\subsection{Gravitational lensing}\label{sec:glens}
We also study the effect of the brightness magnification due to the gravitational lensing. Even if we observe a random field for the survey, the SN brightness could be amplified because of the galaxies happened to exist at the line of sight. To study the effect of the gravitational lensing towards the random field, we use the magnification probability distribution obtained by \citet{hilbert2007lensing} based on ray tracing calculations through the Millennium simulation \citep{springel2005}.  The distributions are updated versions that include the effect of baryons \citep{hilbert2008}. As presented in \citet{wong2019}, we find the discrepancy between the magnification probability distributions estimated by the HSC SSP survey data and those estimated by using the simulation data. This discrepancy is likely from the incompleteness in the observational data and, therefore, we adopt the magnification probability distribution estimated based on the simulation.

The probability of the magnification amplification can be significantly increased when the survey field is towards a massive cluster of galaxies. We refer to \citet{wong2019} for the details of the cluster magnification calculations, but provide a brief summary here. We calculate the magnification distribution along lines of sight towards seven known massive ($\mathrm{M} > 10^{15} \mathrm{M_{\odot}}$) clusters of galaxies to estimate this effect and extrapolate it to our mock survey.  These seven cluster models include the model of the massive cluster J0850+3604 from \citet{wong2017}, as well as models of the six {\it HST} Frontier Fields \citep{lotz2017} clusters from \citet{kawamata2016,kawamata2018} constructed using \citet{oguri2010lens}. Both methods use parameterized models that account for both the cluster dark matter distribution and individual galaxies.  We assume a $14'$ x $14'$ field-of-view.  We calculate the source plane area as a function of magnification for each of the seven clusters and take the average, extrapolated to the full area of our mock survey, to calculate the expected number of detections.  This is somewhat optimistic, as these seven clusters are already among the most massive ones known, but there are potentially other clusters of similar mass that are relatively unexplored (e.g., \cite{wong2013}), and wide-area imaging surveys such as HSC, LSST, and Euclid could potentially find others.  In regions of the source plane that are multiply-imaged, we take the magnification of the brightest image as the value at that particular location.

\subsection{Simulation code}
We use our own code to conduct mock observations to estimate the PISN detectability. The redshifts beyond 5 are binned with $\Delta z =0.1$ and PISNe are assumed to appear in each redshift bin with the PISN rates estimated in the previous section. Once a PISN appears at a redshift bin, the apparent magnitudes in the observer frame at the time of the observation that is determined by \tint\ and the observational period are evaluated. For this purpose, we use redshifted PISN LCs that are calculated based on the PISN spectral model of \citet{kasen2011pisn} in advance for each redshifts. We check the expected observations and judge if the observations match our criteria, such as the limiting magnitudes and \Nd, to regard them as a discovery. When we take the gravitational lensing effect into account, we randomly change PISN magnitudes based on the assumed magnification probability distribution.

\begin{figure}
 \begin{center}
  \includegraphics[width=8cm]{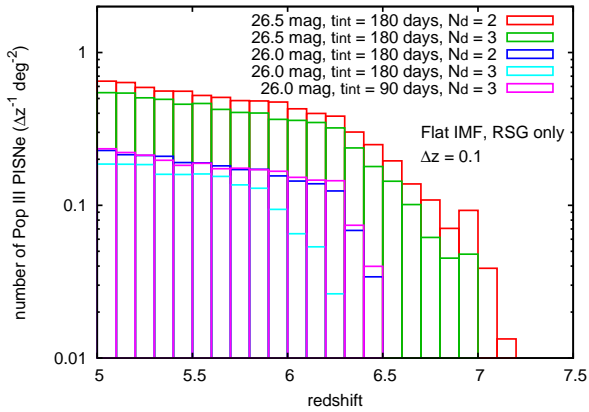} 
 \end{center}
\caption{
Differences in Pop~III PISN discoveries from the different survey parameters with ULTIMATE-Subaru.
}
\label{fig:pisndiffstr}
\end{figure}

\begin{figure}[t]
 \begin{center}
  \includegraphics[width=8cm]{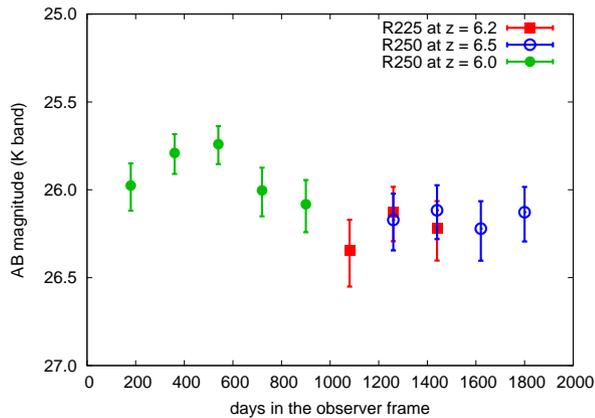} 
 \end{center}
\caption{
Examples of Pop~III PISN LCs observed by the $\tint = 180~\mathrm{days}$ survey with a 26.5~mag limit in the $K$ band with ULTIMATE-Subaru.
}
\label{fig:obslc}
\end{figure}

\begin{figure}
 \begin{center}
  \includegraphics[width=8cm]{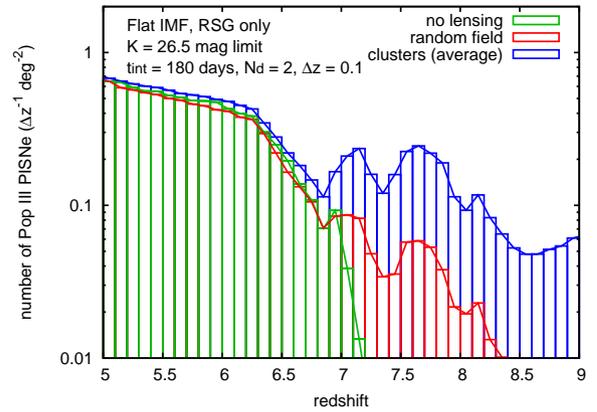} 
 \end{center}
\caption{
Effect of the magnitude amplification due to gravitational lensing.
}\label{fig:lensing}
\end{figure}

\section{Results}\label{sec:results}
Mock transient surveys with one condition are repeated 100 times to check statistical errors. All the errors shown in the following number estimates are the $1\sigma$ statistical errors.

\subsection{ULTIMATE-Subaru}
Table~\ref{tab:pisnnum} summarizes the numbers of Pop~III RSG PISNe discovered in our mock transient surveys with several observational strategies. Fig.~\ref{fig:pisndiffstr} shows the redshift distributions of the observed Pop~III RSG PISNe in the case of the flat IMF. The examples of the observed Pop~III PISN LCs for the survey with $\tint=180~\mathrm{days}$ are presented in Fig.~\ref{fig:obslc}.

We find that $\tint=180~\mathrm{days}$ works well to find Pop~III PISNe at $z\gtrsim6$. If we set the survey FoV to be $1~\mathrm{deg^2}$ in the 26.5~mag surveys ($860$~hours in total with $\tint=180~\mathrm{days}$) and $2~\mathrm{deg^2}$ in the 26.0~mag surveys ($960$~hours in total with $\tint=180~\mathrm{days}$), we predict to find a few Pop~III PISNe at $z\gtrsim 6$ during the 5-year survey if the IMF is flat. If we assume the Salpeter IMF, the expected number of RSG Pop~III PISN detections is as low as $\sim 0.1$ in the whole survey.

The 26.5~mag limit surveys are predicted to find much more Pop~III RSG PISNe because the deeper surveys can find less massive Pop~III PISN progenitors. Comparing the 26.0~mag $2~\mathrm{deg^2}$ survey that takes about 960~hours with the 26.5~mag $1~\mathrm{deg^2}$ survey that takes about 860~hours, having a deeper survey with a smaller FoV is likely beneficial to find Pop~III PISNe than having a shallower survey with a larger FoV (Table~\ref{tab:pisnnum}).

The expected number of Pop~III RSG PISN discoveries can be enhanced if we perform the transient survey towards clusters of galaxies. Table~\ref{tab:lens} and Fig.~\ref{fig:lensing} summarize the expected numbers of Pop~III RSG PISN discoveries obtained by adopting the gravitational amplification probability distributions (see Section~\ref{sec:glens} for details). We focus the $K=26.5~\mathrm{mag}$ survey with $\tint = 180~\mathrm{days}$ and $\Nd = 2$ in this study to present the possibility to reach very high redshifts by the gravitational lensing by clusters of galaxies. We find that the cluster lensing roughly doubles the expected number of Pop~III RSG PISN detections at $z>6$. We also find that the cluster lensing significantly increases the chance to observe PISNe at $z>7$ as seen in Fig.~\ref{fig:lensing}. The bumps in the detection numbers come from the bumps in the Pop~III PISN rates (Fig.~\ref{fig:pisnrate}).

\begin{table}
\tbl{
Numbers of RSG Pop~III PISN discoveries for the ULTIMATE-Subaru 5-year 1-$\mathrm{deg^2}$ $\tint=180~\mathrm{days}$ $\Nd=2$ survey with gravitational lensing.
}{%
\begin{tabular}{lccc}  
\hline\noalign{\vskip3pt} 
source & $z>5$ & $z>6$ & $z>7$  \\ 
\hline\noalign{\vskip3pt} 
&\multicolumn{3}{c}{$K=26.5$ mag limit} \\
no lensing &  $7.9\pm0.4$ & $2.4\pm0.2$ & $0.06\pm0.03$ \\
random field & $8.1\pm0.4$ & $2.9\pm0.2$ & $0.7\pm0.1$ \\
cluster average & $12.2 \pm 1.7$ & $6.4 \pm 1.5 $ & $3.6\pm 1.3$ \\
\hline
&\multicolumn{3}{c}{$K=26.0$ mag limit} \\
no lensing & $2.4 \pm 0.2$ & $0.52\pm 0.08$ & $0$ \\
random field & $2.9\pm0.2$ & $0.61\pm0.09$ & $0.08\pm0.03$ \\
cluster average & $5.5 \pm 1.1$ & $2.2\pm 0.8$ & $1.1 \pm 0.6 $ \\
\hline\noalign{\vskip3pt} 
\end{tabular}}\label{tab:lens}
\end{table}

\subsection{WFIRST}
\begin{figure}[t]
 \begin{center}
  \includegraphics[width=8cm]{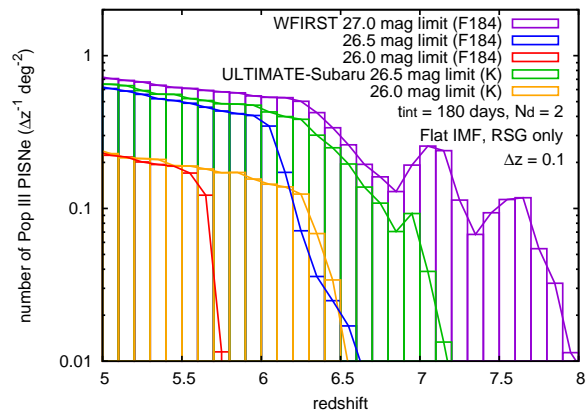} 
 \end{center}
\caption{
Results of the WFIRST Pop~III RSG PISN mock surveys. The numbers are per $\mathrm{deg^2}$ and the actual discovery numbers are proportional to the survey area.
}\label{fig:wfirst}
\end{figure}

Table~\ref{tab:pisnnum} presents the WFIRST Pop~III RSG PISN discoveries and Figure~\ref{fig:wfirst} shows their redshift distributions. Although WFIRST does not reach as red as ULTIMATE-Subaru, it can easily reach deeper limiting magnitudes thanks to being in space. We find that the $F184$-band survey with the limiting magnitude of 26.5~mag with WFIRST can discover PISNe as distant as the $K$-band survey with the limiting magnitude of 26.0~mag with ULTIMATE-Subaru can do is. If we compare the same limiting magnitude survey with WFIRST and ULTIMATE-Subaru, ULTIMATE-Subaru can reach higher redshifts thanks to its redder band. However, WFIRST can reach deeper limiting magnitudes rather easily and the WFIRST survey with the limiting magnitude of 27.0~mag can go beyond the ULTIMATE-Subaru survey with the limiting magnitude of 26.5~mag.

\section{Discussion}\label{sec:discussion}
\subsection{Helium core PISNe}\label{sec:heliumcore}
Pop~III PISN discovery estimates in the previous section do not take helium core PISN progenitors into account. One possible path for helium core Pop~III PISNe to appear is by stripping the envelope of massive RSG PISN progenitors through, e.g., binary interaction. The helium core mass of the R250 model is 124~\Msun\ \citep{kasen2011pisn}. Because helium core PISNe with the helium core mass above $\simeq 125~\Msun$ is able to observe in our surveys (Section~\ref{sec:rates}), Pop~III massive stars with the ZAMS masses of $\sim 250~\Msun$, which are observable if they explode as RSGs, may not be observable if they explode as helium stars. Thus, the expected detection number could go down if some massive RSG Pop~III PISN progenitors explode as bare helium core Pop~III PISNe. 

On the other hand, bare helium core Pop~III PISN progenitors can originate from massive stars with relatively low ZAMS mass (e.g., \cite{chatzopoulos2012pisnrot}). For example, if massive stars rotate rapidly, the rapid rotations can enhance the internal chemical mixing in massive stars and they could evolve chemically homogeneously \citep{yoon2005chemhgrb}. In this case, the ZAMS mass of the He130 progenitor could be as low as $\sim 130~\Msun$. Interestingly enough, a large fraction of Pop~III massive stars might be rapidly rotating stars \citep{hirano2018popiirot}. Close binary systems could also make massive stars to evolve through the chemically homogeneous channel (e.g., \cite{marchant2016ligobh}).  It is also possible that rapidly rotating massive stars originate from mergers of two low mass stars (e.g., \cite{vandenhuevel2013cl}).

In summary, it is currently very hard to quantify how the bare helium core PISN channel would affect the PISN detectability. We speculate that the expected RSG Pop~III discovery number we discussed in the previous section could be lower limits because of the many possible ways to make relatively low mass massive stars to be massive enough to explode as bare helium core PISN progenitors. However, we speculate that the $K$~band survey with ULTIMATE-Subaru is much better to detect the helium core PISNe because of their faintness in the $F184$ band (Fig.~\ref{fig:pisnlcs}). In this sense, ULTIMATE-Subaru has an advantage to find different kinds of Pop~III PISNe.

\subsection{Pop~II PISNe}
We have focused on Pop~III PISN discoveries in this study. At $z\sim6$, however, Pop~II SFR is expected to be $\sim 100$ higher than Pop~III SFR (e.g., \cite{wise2012earlysfr}). On the other hand, Pop~II IMF might be closer to the Salpeter IMF than the flat IMF we mainly focused in our Pop~III study. If Pop~II IMF is approximated as the Salpeter IMF and Pop~II PISN properties are similar to Pop~III PISN properties, we would roughly expect about 100 times more PISNe than the Pop~III PISN number estimates for the Salpeter IMF. This means that we typically expect to detect about $10$ or more Pop~II PISNe at $z\sim 6$ in our deepest surveys with ULTIMATE-Subaru and WFIRST if Pop~II PISNe follows the Salpeter IMF (see Table~\ref{tab:pisnnum}). Thus, in addition to several Pop~III PISNe, we might be able to find many Pop~II PISNe in the deep NIR transient surveys with ULTIMATE-Subaru and WFIRST.

\subsection{SLSNe}
NIR transient surveys discussed in this paper not only have capabilities to find Pop~III PISNe but also high-redshift SLSNe. Detectability of high-redshift SLSNe with WFIRST was estimated in \citet{tanaka2013nirslsndetec}. By scaling their results, we expect to detect $\sim 10$ SLSNe at $z\gtrsim 1$, including a few SLSNe at $z\gtrsim 6$, with a $1~\mathrm{deg^{2}}$ and 26~mag limit survey in NIR. A similar number is expected for a $1~\mathrm{deg^{2}}$ and 26.5~mag limit survey survey.

\subsection{Identifying Pop~III PISNe}
We have shown that discovering Pop~III PISNe is possible by performing the proper transient surveys using ULTIMATE-Subaru and WFIRST. However, in order to confirm the Pop~III PISN discovery, it is necessary to follow up the PISN candidates. In the era of ULTIMATE-Subaru and WFIRST operations, we expect that James Webb Space Telescope (JWST) and several 30-m class telescopes such as Thirty Meter Telescope (TMT) are under operation and they will be essential tools for the spectroscopic follow up. 

Even with these follow-up facilities, it is important to consider how to select the Pop~III PISN candidates to follow. The LCs obtained during the survey are already very important information but having several other information is also important. Especially, we have assumed the NIR transient surveys in a single filter here and we do not have color information to select good Pop~III PISN candidates. Performing the NIR transient surveys in a few filters are the best option, but it is also likely that we can perform the transient survey only in a single filter because of the limited telescope time. In this respect, conducting a coordinated simultaneous observational campaign with the same field with ULTIMATE-Subaru and WFIRST is the best option to have both the $K$ band and $F184$ band information. It is also helpful to have simultaneous optical observations in the same field for the efficient candidate selection, because high-redshift SNe should be faint in optical.
These photometric information can be processed by using SN photometric classification methods to search for genuine PISNe. Recently, SN photometric classification schemes have been developing quickly to prepare for the coming era of extensive time domain surveys with, e.g., Large Synoptic Survey Telescope\footnote{\url{https://www.lsst.org/}} (e.g., \cite{ishida2019,ishida2013,charnock2017,moller2019}). They will be essential to reject many contaminants such as SNe~Ia and core-collapse SNe at low redshifts. They can also be trained by using the PISN LC models to directly identify the high-redshift PISN candidates. Having multi-band information including both optical and NIR is also helpful when we adopt photometric classification methods.

Another important information is provided by the host galaxies. If we perform a transient survey with legacy data from many wavelengths such as the COSMOS field, we can use the preexisting host galaxy information to estimate their photometric redshifts, for example. The host galaxy photometric redshifts are found to be useful in finding high-redshift SNe (e.g., \cite{moriya2018shizuca,curtin2018shizuca}). In order to identify $z\gtrsim 6$ galaxies photometrically, we can look for "dropout" galaxies (e.g., \cite{steidel1999dropout,ono2018dropout}). Galaxies at $z\sim 6$ are observed as "$i$-dropout" galaxies and those at $z\sim 7-9$ are observed as "$z$-dropout" or "$Y$-dropout" galaxies. Deep NIR images are required in advance for the identification of these dropout galaxies.

\section{Conclusions}\label{sec:conclusions}
We have performed mock NIR transient surveys with ULTIMATE-Subaru and WFIRST to estimate their expected numbers of Pop~III PISN detections at $z\gtrsim 6$. We adopt Pop~III PISN rates estimated based on the cosmological simulations by \citet{desouza2014pisnrate} and used the Pop~III PISN LC models by \citet{kasen2011pisn}. We found that a few Pop~III PISNe at $z\gtrsim 6$ may be detected if we perform the $1~\mathrm{deg^2}$ $K$-band ULTIMATE-Subaru transient survey for 5~years with the limiting magnitude of 26.5~mag (860~hours in total), assuming the flat IMF for Pop~III stars. If we assume the Salpeter IMF, the expected number is decreased by a factor of 10. If we set the limiting magnitude to be 26.0~mag, we expect about 1 Pop~III PISN detection with the $2~\mathrm{deg^2}$ survey (960~hours in total). We found that the expected numbers of the Pop~III discovery will be doubled if the transient surveys are conducted towards clusters of galaxies thanks to the magnification by the gravitational lensing.

The reddest filter of WFIRST ($F184$) is bluer than that of ULTIMATE-Subaru ($K$). However, WFIRST has the lager FoV than ULTIMATE-Subaru and it allows to conduct wider transient surveys. If we conduct the 5-year transient survey with the $F184$ filter with the limiting magnitude of 26.5~mag for $10~\mathrm{deg^2}$ (200~hours), we expect to find about 7 Pop~III PISNe with the flat IMF and about 0.3 Pop~III PISNe with the Salpeter IMF.

Our study has shown that the deep and wide NIR transient surveys conducted by the planned wide-field NIR imagers will enable us to acquire valuable information on the first generation stars in the Universe. They have a possibility to find Pop~III PISNe. Even if we do not discovery any Pop~III PISNe, such NIR transient surveys will enable us to constrain the star-formation properties like IMF of the first stars.





\begin{ack}
TJM thanks horrible weather at Mauna Kea in Feb and Mar 2018 which led to the cancellation of his Subaru observations that made this study done. We thank Dan Kasen for sharing electric data of PISN LC models. TJM thanks Chien-Hsiu Lee for discussion.
TJM is supported by the Grants-in-Aid for Scientific Research of the Japan Society for the Promotion of Science (16H07413, 17H02864, 18K13585).
KCW is supported in part
by an EACOA Fellowship awarded by the East Asia Core Observatories
Association, which consists of the Academia Sinica Institute of
Astronomy and Astrophysics, the National Astronomical Observatory of
Japan, the National Astronomical Observatories of the Chinese Academy
of Sciences, and the Korea Astronomy and Space Science Institute.
MO is supported in part by JSPS KAKENHI Grant Number JP15H05892 and JP18K03693.

This work was supported by World Premier International Research Center
Initiative (WPI Initiative), MEXT, Japan.  

The Hyper Suprime-Cam (HSC) collaboration includes the astronomical communities of Japan and Taiwan, and Princeton University.  The HSC instrumentation and software were developed by the National Astronomical Observatory of Japan (NAOJ), the Kavli Institute for the Physics and Mathematics of the Universe (Kavli IPMU), the University of Tokyo, the High Energy Accelerator Research Organization (KEK), the Academia Sinica Institute for Astronomy and Astrophysics in Taiwan (ASIAA), and Princeton University.  Funding was contributed by the FIRST program from Japanese Cabinet Office, the Ministry of Education, Culture, Sports, Science and Technology (MEXT), the Japan Society for the Promotion of Science (JSPS), Japan Science and Technology Agency (JST), the Toray Science Foundation, NAOJ, Kavli IPMU, KEK, ASIAA, and Princeton University.

The Pan-STARRS1 Surveys (PS1) have been made possible through contributions of the Institute for Astronomy, the University of Hawaii, the Pan-STARRS Project Office, the Max-Planck Society and its participating institutes, the Max Planck Institute for Astronomy, Heidelberg and the Max Planck Institute for Extraterrestrial Physics, Garching, The Johns Hopkins University, Durham University, the University of Edinburgh, Queen's University Belfast, the Harvard-Smithsonian Center for Astrophysics, the Las Cumbres Observatory Global Telescope Network Incorporated, the National Central University of Taiwan, the Space Telescope Science Institute, the National Aeronautics and Space Administration under Grant No. NNX08AR22G issued through the Planetary Science Division of the NASA Science Mission Directorate, the National Science Foundation under Grant No. AST-1238877, the University of Maryland, and Eotvos Lorand University (ELTE).

This paper makes use of software developed for the Large Synoptic Survey Telescope. We thank the LSST Project for making their code available as free software at \url{http://dm.lsst.org}.

Based in part on data collected at the Subaru Telescope and retrieved from the HSC data archive system, which is operated by the Subaru Telescope and Astronomy Data Center at National Astronomical Observatory of Japan.
\end{ack}

\bibliographystyle{myaasjournal}
\bibliography{references}

\begin{thebibliography}{}
\expandafter\ifx\csname natexlab\endcsname\relax\def\natexlab#1{#1}\fi
\providecommand{\url}[1]{\href{#1}{#1}}

\bibitem[{{Barkat} {et~al.}(1967){Barkat}, {Rakavy}, \&
  {Sack}}]{barkat1967pisn}
{Barkat}, Z., {Rakavy}, G., \& {Sack}, N. 1967, Physical Review Letters, 18,
  379

\bibitem[{{Chambers} {et~al.}(2016){Chambers}, {Magnier}, {Metcalfe},
  {Flewelling}, {Huber}, {Waters}, {Denneau}, {Draper}, {Farrow}, {Finkbeiner},
  {Holmberg}, {Koppenhoefer}, {Price}, {Saglia}, {Schlafly}, {Smartt},
  {Sweeney}, {Wainscoat}, {Burgett}, {Grav}, {Heasley}, {Hodapp}, {Jedicke},
  {Kaiser}, {Kudritzki}, {Luppino}, {Lupton}, {Monet}, {Morgan}, {Onaka},
  {Stubbs}, {Tonry}, {Banados}, {Bell}, {Bender}, {Bernard}, {Botticella},
  {Casertano}, {Chastel}, {Chen}, {Chen}, {Cole}, {Deacon}, {Frenk},
  {Fitzsimmons}, {Gezari}, {Goessl}, {Goggia}, {Goldman}, {Grebel}, {Hambly},
  {Hasinger}, {Heavens}, {Heckman}, {Henderson}, {Henning}, {Holman}, {Hopp},
  {Ip}, {Isani}, {Keyes}, {Koekemoer}, {Kotak}, {Long}, {Lucey}, {Liu},
  {Martin}, {McLean}, {Morganson}, {Murphy}, {Nieto-Santisteban}, {Norberg},
  {Peacock}, {Pier}, {Postman}, {Primak}, {Rae}, {Rest}, {Riess}, {Riffeser},
  {Rix}, {Roser}, {Schilbach}, {Schultz}, {Scolnic}, {Szalay}, {Seitz},
  {Shiao}, {Small}, {Smith}, {Soderblom}, {Taylor}, {Thakar}, {Thiel},
  {Thilker}, {Urata}, {Valenti}, {Walter}, {Watters}, {Werner}, {White},
  {Wood-Vasey}, \& {Wyse}}]{chambers2016panstarrs}
{Chambers}, K.~C., {et~al.} 2016, ArXiv e-prints, arXiv:1612.05560

\bibitem[{{Charnock} \& {Moss}(2017)}]{charnock2017}
{Charnock}, T., \& {Moss}, A. 2017, \apjl, 837, L28

\bibitem[{{Chatzopoulos} \& {Wheeler}(2012)}]{chatzopoulos2012pisnrot}
{Chatzopoulos}, E., \& {Wheeler}, J.~C. 2012, \apj, 748, 42

\bibitem[{{Curtin} {et~al.}(2019){Curtin}, {Cooke}, {Moriya}, {Bernard},
  {Galbany}, {Jiang}, {Lee}, {Maeda}, {Morokuma}, {Nomoto}, {Pignata},
  {Pritchard}, {Quimby}, {Suzuki}, {Takahashi}, {Tanaka}, {Tanaka}, {Tominaga},
  {Yamaguchi}, \& {Yasuda}}]{curtin2018shizuca}
{Curtin}, C., {et~al.} 2019, ArXiv e-prints, arXiv:1801.08241

\bibitem[{{de Souza} {et~al.}(2013){de Souza}, {Ishida}, {Johnson}, {Whalen},
  \& {Mesinger}}]{desouza2013jwst}
{de Souza}, R.~S., {Ishida}, E.~E.~O., {Johnson}, J.~L., {Whalen}, D.~J., \&
  {Mesinger}, A. 2013, \mnras, 436, 1555

\bibitem[{{de Souza} {et~al.}(2014){de Souza}, {Ishida}, {Whalen}, {Johnson},
  \& {Ferrara}}]{desouza2014pisnrate}
{de Souza}, R.~S., {Ishida}, E.~E.~O., {Whalen}, D.~J., {Johnson}, J.~L., \&
  {Ferrara}, A. 2014, \mnras, 442, 1640

\bibitem[{{Dessart} {et~al.}(2012){Dessart}, {Hillier}, {Waldman}, {Livne}, \&
  {Blondin}}]{dessart2012magslsn}
{Dessart}, L., {Hillier}, D.~J., {Waldman}, R., {Livne}, E., \& {Blondin}, S.
  2012, \mnras, 426, L76

\bibitem[{{Gal-Yam} {et~al.}(2009){Gal-Yam}, {Mazzali}, {Ofek}, {Nugent},
  {Kulkarni}, {Kasliwal}, {Quimby}, {Filippenko}, {Cenko}, {Chornock},
  {Waldman}, {Kasen}, {Sullivan}, {Beshore}, {Drake}, {Thomas}, {Bloom},
  {Poznanski}, {Miller}, {Foley}, {Silverman}, {Arcavi}, {Ellis}, \&
  {Deng}}]{gal-yam2009sn2007bi}
{Gal-Yam}, A., {et~al.} 2009, \nat, 462, 624

\bibitem[{{Georgy} {et~al.}(2017){Georgy}, {Meynet}, {Ekstr{\"o}m}, {Wade},
  {Petit}, {Keszthelyi}, \& {Hirschi}}]{georgy2017solarpisn}
{Georgy}, C., {Meynet}, G., {Ekstr{\"o}m}, S., {Wade}, G.~A., {Petit}, V.,
  {Keszthelyi}, Z., \& {Hirschi}, R. 2017, \aap, 599, L5

\bibitem[{{Hartwig} {et~al.}(2018){Hartwig}, {Bromm}, \&
  {Loeb}}]{hartwig2018jwstpisn}
{Hartwig}, T., {Bromm}, V., \& {Loeb}, A. 2018, \mnras, arXiv:1711.05742

\bibitem[{{Heger} \& {Woosley}(2002)}]{heger2002popiii}
{Heger}, A., \& {Woosley}, S.~E. 2002, \apj, 567, 532

\bibitem[{{Hilbert} {et~al.}(2007){Hilbert}, {White}, {Hartlap}, \&
  {Schneider}}]{hilbert2007lensing}
{Hilbert}, S., {White}, S.~D.~M., {Hartlap}, J., \& {Schneider}, P. 2007,
  \mnras, 382, 121

\bibitem[{{Hilbert} {et~al.}(2008){Hilbert}, {White}, {Hartlap}, \&
  {Schneider}}]{hilbert2008}
---. 2008, \mnras, 386, 1845

\bibitem[{{Hirano} \& {Bromm}(2018)}]{hirano2018popiirot}
{Hirano}, S., \& {Bromm}, V. 2018, \mnras, 476, 3964

\bibitem[{{Hirano} {et~al.}(2015){Hirano}, {Hosokawa}, {Yoshida}, {Omukai}, \&
  {Yorke}}]{hirano2015popiii}
{Hirano}, S., {Hosokawa}, T., {Yoshida}, N., {Omukai}, K., \& {Yorke}, H.~W.
  2015, \mnras, 448, 568

\bibitem[{{Hounsell} {et~al.}(2017){Hounsell}, {Scolnic}, {Foley}, {Kessler},
  {Miranda}, {Avelino}, {Bohlin}, {Filippenko}, {Frieman}, {Jha}, {Kelly},
  {Kirshner}, {Mandel}, {Rest}, {Riess}, {Rodney}, \&
  {Strolger}}]{hounsell2017wfirstia}
{Hounsell}, R., {et~al.} 2017, ArXiv e-prints, arXiv:1702.01747

\bibitem[{{Ichikawa} {et~al.}(2006){Ichikawa}, {Suzuki}, {Tokoku}, {Uchimoto},
  {Konishi}, {Yoshikawa}, {Yamada}, {Tanaka}, {Omata}, \&
  {Nishimura}}]{ichikawa2006moircs}
{Ichikawa}, T., {et~al.} 2006, in \procspie, Vol. 6269, Society of
  Photo-Optical Instrumentation Engineers (SPIE) Conference Series, 626916

\bibitem[{{Inserra} {et~al.}(2017){Inserra}, {Nichol}, {Scovacricchi},
  {Amiaux}, {Brescia}, {Burigana}, {Cappellaro}, {Carvalho}, {Cavuoti},
  {Conforti}, {Cuillandre}, {da Silva}, {De Rosa}, {Della Valle}, {Dinis},
  {Franceschi}, {Hook}, {Hudelot}, {Jahnke}, {Kitching}, {Kurki-Suonio},
  {Lloro}, {Longo}, {Maiorano}, {Maris}, {Rhodes}, {Scaramella}, {Smartt},
  {Sullivan}, {Tao}, {Toledo-Moreo}, {Tereno}, {Trifoglio}, \&
  {Valenziano}}]{inserra2017euclid}
{Inserra}, C., {et~al.} 2017, ArXiv e-prints, arXiv:1710.09585

\bibitem[{{Ishida} \& {de Souza}(2013)}]{ishida2013}
{Ishida}, E.~E.~O., \& {de Souza}, R.~S. 2013, \mnras, 430, 509

\bibitem[{{Ishida} {et~al.}(2019){Ishida}, {Beck}, {Gonz{\'a}lez-Gait{\'a}n},
  {de Souza}, {Krone-Martins}, {Barrett}, {Kennamer}, {Vilalta}, {Burgess},
  {Quint}, {Vitorelli}, {Mahabal}, \& {Gangler}}]{ishida2019}
{Ishida}, E.~E.~O., {et~al.} 2019, \mnras, 483, 2

\bibitem[{{Jerkstrand} {et~al.}(2016){Jerkstrand}, {Smartt}, \&
  {Heger}}]{jerkstrand2016pisnnebular}
{Jerkstrand}, A., {Smartt}, S.~J., \& {Heger}, A. 2016, \mnras, 455, 3207

\bibitem[{{Johnson} {et~al.}(2013){Johnson}, {Dalla Vecchia}, \&
  {Khochfar}}]{johnson2013cosmosim}
{Johnson}, J.~L., {Dalla Vecchia}, C., \& {Khochfar}, S. 2013, \mnras, 428,
  1857

\bibitem[{{Kasen} {et~al.}(2011){Kasen}, {Woosley}, \& {Heger}}]{kasen2011pisn}
{Kasen}, D., {Woosley}, S.~E., \& {Heger}, A. 2011, \apj, 734, 102

\bibitem[{{Kasliwal} {et~al.}(2017){Kasliwal}, {Bally}, {Masci}, {Cody},
  {Bond}, {Jencson}, {Tinyanont}, {Cao}, {Contreras}, {Dykhoff}, {Amodeo},
  {Armus}, {Boyer}, {Cantiello}, {Carlon}, {Cass}, {Cook}, {Corgan}, {Faella},
  {Fox}, {Green}, {Gehrz}, {Helou}, {Hsiao}, {Johansson}, {Khan}, {Lau},
  {Langer}, {Levesque}, {Milne}, {Mohamed}, {Morrell}, {Monson}, {Moore},
  {Ofek}, {O' Sullivan}, {Parthasarathy}, {Perez}, {Perley}, {Phillips},
  {Prince}, {Shenoy}, {Smith}, {Surace}, {Van Dyk}, {Whitelock}, \&
  {Williams}}]{kasliwal2017spirits}
{Kasliwal}, M.~M., {et~al.} 2017, \apj, 839, 88

\bibitem[{{Kawamata} {et~al.}(2018){Kawamata}, {Ishigaki}, {Shimasaku},
  {Oguri}, {Ouchi}, \& {Tanigawa}}]{kawamata2018}
{Kawamata}, R., {Ishigaki}, M., {Shimasaku}, K., {Oguri}, M., {Ouchi}, M., \&
  {Tanigawa}, S. 2018, \apj, 855, 4

\bibitem[{{Kawamata} {et~al.}(2016){Kawamata}, {Oguri}, {Ishigaki},
  {Shimasaku}, \& {Ouchi}}]{kawamata2016}
{Kawamata}, R., {Oguri}, M., {Ishigaki}, M., {Shimasaku}, K., \& {Ouchi}, M.
  2016, \apj, 819, 114

\bibitem[{{Keller} {et~al.}(2007){Keller}, {Schmidt}, {Bessell}, {Conroy},
  {Francis}, {Granlund}, {Kowald}, {Oates}, {Martin-Jones}, {Preston},
  {Tisserand}, {Vaccarella}, \& {Waterson}}]{keller2007skymapper}
{Keller}, S.~C., {et~al.} 2007, PASA, 24, 1

\bibitem[{{Kool} {et~al.}(2018){Kool}, {Ryder}, {Kankare}, {Mattila},
  {Reynolds}, {McDermid}, {P{\'e}rez-Torres}, {Herrero-Illana}, {Schirmer},
  {Efstathiou}, {Bauer}, {Kotilainen}, {V{\"a}is{\"a}nen}, {Baldwin},
  {Romero-Ca{\~n}izales}, \& {Alberdi}}]{kool2018sunbird}
{Kool}, E.~C., {et~al.} 2018, \mnras, 473, 5641

\bibitem[{{Kozyreva} {et~al.}(2016){Kozyreva}, {Gilmer}, {Hirschi},
  {Fr{\"o}hlich}, {Blinnikov}, {Wollaeger}, {Noebauer}, {van Rossum}, {Heger},
  {Even}, {Waldman}, {Tolstov}, {Chatzopoulos}, \&
  {Sorokina}}]{kozyreva2016rapidpisn}
{Kozyreva}, A., {et~al.} 2016, \mnras, arXiv:1610.01086

\bibitem[{{Langer}(2012)}]{langer2012review}
{Langer}, N. 2012, \araa, 50, 107

\bibitem[{{Langer} {et~al.}(2007){Langer}, {Norman}, {de Koter}, {Vink},
  {Cantiello}, \& {Yoon}}]{langer2007pisn}
{Langer}, N., {Norman}, C.~A., {de Koter}, A., {Vink}, J.~S., {Cantiello}, M.,
  \& {Yoon}, S.-C. 2007, \aap, 475, L19

\bibitem[{{Law} {et~al.}(2009){Law}, {Kulkarni}, {Dekany}, {Ofek}, {Quimby},
  {Nugent}, {Surace}, {Grillmair}, {Bloom}, {Kasliwal}, {Bildsten}, {Brown},
  {Cenko}, {Ciardi}, {Croner}, {Djorgovski}, {van Eyken}, {Filippenko}, {Fox},
  {Gal-Yam}, {Hale}, {Hamam}, {Helou}, {Henning}, {Howell}, {Jacobsen},
  {Laher}, {Mattingly}, {McKenna}, {Pickles}, {Poznanski}, {Rahmer}, {Rau},
  {Rosing}, {Shara}, {Smith}, {Starr}, {Sullivan}, {Velur}, {Walters}, \&
  {Zolkower}}]{law2009ptf}
{Law}, N.~M., {et~al.} 2009, \pasp, 121, 1395

\bibitem[{{Lotz} {et~al.}(2017){Lotz}, {Koekemoer}, {Coe}, {Grogin}, {Capak},
  {Mack}, {Anderson}, {Avila}, {Barker}, {Borncamp}, {Brammer}, {Durbin},
  {Gunning}, {Hilbert}, {Jenkner}, {Khandrika}, {Levay}, {Lucas}, {MacKenty},
  {Ogaz}, {Porterfield}, {Reid}, {Robberto}, {Royle}, {Smith},
  {Storrie-Lombardi}, {Sunnquist}, {Surace}, {Taylor}, {Williams}, {Bullock},
  {Dickinson}, {Finkelstein}, {Natarajan}, {Richard}, {Robertson}, {Tumlinson},
  {Zitrin}, {Flanagan}, {Sembach}, {Soifer}, \& {Mountain}}]{lotz2017}
{Lotz}, J.~M., {et~al.} 2017, \apj, 837, 97

\bibitem[{{Marchant} {et~al.}(2016){Marchant}, {Langer}, {Podsiadlowski},
  {Tauris}, \& {Moriya}}]{marchant2016ligobh}
{Marchant}, P., {Langer}, N., {Podsiadlowski}, P., {Tauris}, T.~M., \&
  {Moriya}, T.~J. 2016, \aap, 588, A50

\bibitem[{{Mattila} {et~al.}(2012){Mattila}, {Dahlen}, {Efstathiou}, {Kankare},
  {Melinder}, {Alonso-Herrero}, {P{\'e}rez-Torres}, {Ryder},
  {V{\"a}is{\"a}nen}, \& {{\"O}stlin}}]{mattila2012nirsn}
{Mattila}, S., {et~al.} 2012, \apj, 756, 111

\bibitem[{{M{\"o}ller} \& {de Boissi{\`e}re}(2019)}]{moller2019}
{M{\"o}ller}, A., \& {de Boissi{\`e}re}, T. 2019, arXiv e-prints,
  arXiv:1901.06384

\bibitem[{{Moriya} \& {Langer}(2015)}]{moriya2015langer}
{Moriya}, T.~J., \& {Langer}, N. 2015, \aap, 573, A18

\bibitem[{{Moriya} {et~al.}(2018){Moriya}, {Sorokina}, \&
  {Chevalier}}]{moriya2018slsnreview}
{Moriya}, T.~J., {Sorokina}, E.~I., \& {Chevalier}, R.~A. 2018, \ssr, 214, 59

\bibitem[{{Moriya} {et~al.}(2019){Moriya}, {Tanaka}, {Yasuda}, {Jiang}, {Lee},
  {Maeda}, {Morokuma}, {Nomoto}, {Quimby}, {Suzuki}, {Takahashi}, {Tanaka},
  {Tominaga}, {Yamaguchi}, {Bernard}, {Cooke}, {Curtin}, {Galbany},
  {Gonzalez-Gaitan}, {Pignata}, {Pritchard}, {Komiyama}, \&
  {Lupton}}]{moriya2018shizuca}
{Moriya}, T.~J., {et~al.} 2019, ArXiv e-prints, arXiv:1801.08240

\bibitem[{{Morokuma} {et~al.}(2014){Morokuma}, {Tominaga}, {Tanaka}, {Mori},
  {Matsumoto}, {Kikuchi}, {Shibata}, {Sako}, {Aoki}, {Doi}, {Kobayashi},
  {Maehara}, {Matsunaga}, {Mito}, {Miyata}, {Nakada}, {Soyano}, {Tarusawa},
  {Miyazaki}, {Nakata}, {Okada}, {Sarugaku}, {Richmond}, {Akitaya}, {Aldering},
  {Arimatsu}, {Contreras}, {Horiuchi}, {Hsiao}, {Itoh}, {Iwata}, {Kawabata},
  {Kawai}, {Kitagawa}, {Kokubo}, {Kuroda}, {Mazzali}, {Misawa}, {Moritani},
  {Morrell}, {Okamoto}, {Pavlyuk}, {Phillips}, {Pian}, {Sahu}, {Saito}, {Sano},
  {Stritzinger}, {Tachibana}, {Taddia}, {Takaki}, {Tateuchi}, {Tomita},
  {Tsvetkov}, {Ui}, {Ukita}, {Urata}, {Walker}, \& {Yoshii}}]{morokuma2014kiss}
{Morokuma}, T., {et~al.} 2014, \pasj, 66, 114

\bibitem[{{Oguri}(2010)}]{oguri2010lens}
{Oguri}, M. 2010, \pasj, 62, 1017

\bibitem[{{Ono} {et~al.}(2018){Ono}, {Ouchi}, {Harikane}, {Toshikawa}, {Rauch},
  {Yuma}, {Sawicki}, {Shibuya}, {Shimasaku}, {Oguri}, {Willott}, {Akhlaghi},
  {Akiyama}, {Coupon}, {Kashikawa}, {Komiyama}, {Konno}, {Lin}, {Matsuoka},
  {Miyazaki}, {Nagao}, {Nakajima}, {Silverman}, {Tanaka}, {Taniguchi}, \&
  {Wang}}]{ono2018dropout}
{Ono}, Y., {et~al.} 2018, \pasj, 70, S10

\bibitem[{{Quimby} {et~al.}(2011){Quimby}, {Kulkarni}, {Kasliwal}, {Gal-Yam},
  {Arcavi}, {Sullivan}, {Nugent}, {Thomas}, {Howell}, {Nakar}, {Bildsten},
  {Theissen}, {Law}, {Dekany}, {Rahmer}, {Hale}, {Smith}, {Ofek}, {Zolkower},
  {Velur}, {Walters}, {Henning}, {Bui}, {McKenna}, {Poznanski}, {Cenko}, \&
  {Levitan}}]{quimby2011slsn}
{Quimby}, R.~M., {et~al.} 2011, \nat, 474, 487

\bibitem[{{Rakavy} \& {Shaviv}(1967)}]{rakavy1967pisn}
{Rakavy}, G., \& {Shaviv}, G. 1967, \apj, 148, 803

\bibitem[{{Scannapieco} {et~al.}(2005){Scannapieco}, {Madau}, {Woosley},
  {Heger}, \& {Ferrara}}]{scannapieco2005pisn}
{Scannapieco}, E., {Madau}, P., {Woosley}, S., {Heger}, A., \& {Ferrara}, A.
  2005, \apj, 633, 1031

\bibitem[{{Smith} {et~al.}(2007){Smith}, {Li}, {Foley}, {Wheeler}, {Pooley},
  {Chornock}, {Filippenko}, {Silverman}, {Quimby}, {Bloom}, \&
  {Hansen}}]{smith2007sn2006gyearly}
{Smith}, N., {et~al.} 2007, \apj, 666, 1116

\bibitem[{{Spergel} {et~al.}(2015){Spergel}, {Gehrels}, {Baltay}, {Bennett},
  {Breckinridge}, {Donahue}, {Dressler}, {Gaudi}, {Greene}, {Guyon}, {Hirata},
  {Kalirai}, {Kasdin}, {Macintosh}, {Moos}, {Perlmutter}, {Postman},
  {Rauscher}, {Rhodes}, {Wang}, {Weinberg}, {Benford}, {Hudson}, {Jeong},
  {Mellier}, {Traub}, {Yamada}, {Capak}, {Colbert}, {Masters}, {Penny},
  {Savransky}, {Stern}, {Zimmerman}, {Barry}, {Bartusek}, {Carpenter}, {Cheng},
  {Content}, {Dekens}, {Demers}, {Grady}, {Jackson}, {Kuan}, {Kruk}, {Melton},
  {Nemati}, {Parvin}, {Poberezhskiy}, {Peddie}, {Ruffa}, {Wallace}, {Whipple},
  {Wollack}, \& {Zhao}}]{spergel2015wfirst}
{Spergel}, D., {et~al.} 2015, ArXiv e-prints, arXiv:1503.03757

\bibitem[{{Springel} {et~al.}(2005){Springel}, {White}, {Jenkins}, {Frenk},
  {Yoshida}, {Gao}, {Navarro}, {Thacker}, {Croton}, {Helly}, {Peacock}, {Cole},
  {Thomas}, {Couchman}, {Evrard}, {Colberg}, \& {Pearce}}]{springel2005}
{Springel}, V., {et~al.} 2005, \nat, 435, 629

\bibitem[{{Steidel} {et~al.}(1999){Steidel}, {Adelberger}, {Giavalisco},
  {Dickinson}, \& {Pettini}}]{steidel1999dropout}
{Steidel}, C.~C., {Adelberger}, K.~L., {Giavalisco}, M., {Dickinson}, M., \&
  {Pettini}, M. 1999, \apj, 519, 1

\bibitem[{{Suzuki} {et~al.}(2008){Suzuki}, {Tokoku}, {Ichikawa}, {Uchimoto},
  {Konishi}, {Yoshikawa}, {Tanaka}, {Yamada}, {Omata}, \&
  {Nishimura}}]{suzuki2008moircs}
{Suzuki}, R., {et~al.} 2008, \pasj, 60, 1347

\bibitem[{{Tanaka} {et~al.}(2013){Tanaka}, {Moriya}, \&
  {Yoshida}}]{tanaka2013nirslsndetec}
{Tanaka}, M., {Moriya}, T.~J., \& {Yoshida}, N. 2013, \mnras, 435, 2483

\bibitem[{{Tanaka} {et~al.}(2016){Tanaka}, {Tominaga}, {Morokuma}, {Yasuda},
  {Furusawa}, {Baklanov}, {Blinnikov}, {Moriya}, {Doi}, {Jiang}, {Kato},
  {Kikuchi}, {Kuncarayakti}, {Nagao}, {Nomoto}, \&
  {Taniguchi}}]{tanaka2016hscrapidrise}
{Tanaka}, M., {et~al.} 2016, \apj, 819, 5

\bibitem[{{Tolstov} {et~al.}(2017){Tolstov}, {Zhiglo}, {Nomoto}, {Sorokina},
  {Kozyreva}, \& {Blinnikov}}]{tolstov2017gaia}
{Tolstov}, A., {Zhiglo}, A., {Nomoto}, K., {Sorokina}, E., {Kozyreva}, A., \&
  {Blinnikov}, S. 2017, \apjl, 845, L2

\bibitem[{{Umeda} \& {Nomoto}(2002)}]{umeda2002popiii}
{Umeda}, H., \& {Nomoto}, K. 2002, \apj, 565, 385

\bibitem[{{van den Heuvel} \& {Portegies Zwart}(2013)}]{vandenhuevel2013cl}
{van den Heuvel}, E.~P.~J., \& {Portegies Zwart}, S.~F. 2013, \apj, 779, 114

\bibitem[{{Whalen} {et~al.}(2013{\natexlab{a}}){Whalen}, {Smidt}, {Rydberg},
  {Johnson}, {Holz}, \& {Stiavelli}}]{whalen2013lensing}
{Whalen}, D.~J., {Smidt}, J., {Rydberg}, C.-E., {Johnson}, J.~L., {Holz},
  D.~E., \& {Stiavelli}, M. 2013{\natexlab{a}}, arXiv e-prints, arXiv:1312.6330

\bibitem[{{Whalen} {et~al.}(2013{\natexlab{b}}){Whalen}, {Even}, {Frey},
  {Smidt}, {Johnson}, {Lovekin}, {Fryer}, {Stiavelli}, {Holz}, {Heger},
  {Woosley}, \& {Hungerford}}]{whalen2013pisn}
{Whalen}, D.~J., {et~al.} 2013{\natexlab{b}}, \apj, 777, 110

\bibitem[{{Whalen} {et~al.}(2014){Whalen}, {Smidt}, {Heger}, {Hirschi},
  {Yusof}, {Even}, {Fryer}, {Stiavelli}, {Chen}, \&
  {Joggerst}}]{whalen2014pisn}
---. 2014, \apj, 797, 9

\bibitem[{{Wise} {et~al.}(2012){Wise}, {Turk}, {Norman}, \&
  {Abel}}]{wise2012earlysfr}
{Wise}, J.~H., {Turk}, M.~J., {Norman}, M.~L., \& {Abel}, T. 2012, \apj, 745,
  50

\bibitem[{{Wong} {et~al.}(2019){Wong}, {Moriya}, {Oguri}, {Hilbert}, \&
  {Koyama}}]{wong2019}
{Wong}, K.~C., {Moriya}, T.~J., {Oguri}, M., {Hilbert}, S., \& {Koyama}, Y.
  2019, \pasj, submitted

\bibitem[{{Wong} {et~al.}(2017){Wong}, {Raney}, {Keeton}, {Umetsu},
  {Zabludoff}, {Ammons}, \& {French}}]{wong2017}
{Wong}, K.~C., {Raney}, C., {Keeton}, C.~R., {Umetsu}, K., {Zabludoff}, A.~I.,
  {Ammons}, S.~M., \& {French}, K.~D. 2017, \apj, 844, 127

\bibitem[{{Wong} {et~al.}(2013){Wong}, {Zabludoff}, {Ammons}, {Keeton}, {Hogg},
  \& {Gonzalez}}]{wong2013}
{Wong}, K.~C., {Zabludoff}, A.~I., {Ammons}, S.~M., {Keeton}, C.~R., {Hogg},
  D.~W., \& {Gonzalez}, A.~H. 2013, \apj, 769, 52

\bibitem[{{Yoon} {et~al.}(2012){Yoon}, {Dierks}, \& {Langer}}]{yoon2012popiii}
{Yoon}, S.-C., {Dierks}, A., \& {Langer}, N. 2012, \aap, 542, A113

\bibitem[{{Yoon} \& {Langer}(2005)}]{yoon2005chemhgrb}
{Yoon}, S.-C., \& {Langer}, N. 2005, \aap, 443, 643

\bibitem[{{Yoshida} {et~al.}(2014){Yoshida}, {Okita}, \&
  {Umeda}}]{yoshida2014massiveic}
{Yoshida}, T., {Okita}, S., \& {Umeda}, H. 2014, \mnras, 438, 3119

\end{thebibliography}

\end{document}